\documentclass{jpp}

\usepackage{graphicx}
\usepackage{natbib}

\usepackage{upmath}
\usepackage{amssymb}
\usepackage{amsbsy}

\usepackage[T2A]{fontenc}
\usepackage{amsmath}
\usepackage{amsfonts}

\newcommand\etal{\mbox{\textit{et al.}}}

\title[Path integrals for mean-field dynamos]{Path integrals for mean-field equations in nonlinear dynamos}

\author[Dmitry Sokoloff and Nobumitsu Yokoi]%
{D\ls M\ls I\ls T\ls R\ls Y\ns  S\ls O\ls K\ls O\ls L\ls O\ls F\ls F $^{1,2}$
  \thanks{Email address for correspondence: sokoloff.dd@gmail.com},\ns
\and \ns N\ls O\ls B\ls U\ls M\ls I\ls T\ls S\ls U \ls Y\ls O\ls K\ls O\ls I$^3$
}

\affiliation{$^1$Department of Physics,
Moscow State University, Moscow, 119991, Russia\\
$^2$IZMIRAN, Kaluzhskoe shosse, Troitsk, Moscow, 108840, Russia \\
$^3$Institute of Industrial Science, University of Tokyo, Tokyo 153-8505, Japan}

\pubyear{2018}
\volume{}
\pagerange{}
\date{?; revised ?; accepted ?.}

\begin{document}

\maketitle

\begin{abstract}
Mean-field dynamo equations are addressed with the aid of the path-integral method. The evolution of magnetic 
field is treated as a three-dimensional Wiener random process, and the mean magnetic-field equations are 
obtained with the Wiener integral over all the trajectories of fluid particle. The form of the equations is 
just the same as the conventional mean-field equations, but the present equations are derived with the 
velocity-field realization affected by the magnetic-field force. In this sense, the present ones are nonlinear 
dynamo equations. 
\end{abstract}

\begin{PACS}
52.30.Cv; 96.12.Hg; 96.25.St; 98.38.-j 
\end{PACS}

\section{Introduction}

Mean-field equations were introduced in dynamo theory at quite early stage of its development (see e.g. \cite{KR}). That time it was the main tool to 
produce dynamo models for magnetic field evolution in various celestial bodies including the Sun.
Contemporary science has for this aim other tools including in the first line  direct numerical simulations 
however mean-field equation remains an important approach in the field. In particular, 
mean-field description is useful to understand physical mechanism of dynamo action based on  mirror 
asymmetry of turbulence or convection in form of famous $\alpha$-effect which explains 
solar dynamo mechanism suggested in \cite{P55}.

Several methods to perform statistical averaging of induction equation in order to obtain mean-field 
dynamo equations 
were suggested. The main issue here is how to split 
correlations in terms which contain both velocity field $\bf v$ as well as magnetic field $\bf H$ which 
are obviously statistically dependent. One of such method is path integral method initially suggested in 
\cite{Metal83}.

The idea of the method can be presented as follows. Performing statistical averaging it looks attractive 
to deal with solution of induction equation rather with induction equation itself because solution by 
definition contains velocity field and initial magnetic field only. 
The problem is how to get such a solution. It is affordable if we neglect Ohmic losses and use Lagrangian 
reference frame so induction equation reduces to a system of linear ordinary differential equations. Then 
the desired solution can be obtained following magnetic field evolution over the Largangian trajectory.

Including Ohmic losses, we have to consider a bundle of random paths surrounding the Lagrandian 
trajectory. These paths arise due to joint action of advection and random walks which mimic 
magnetic field diffusion. Magnetic field evolution is considered here as a superposition of contributions 
associated with particular random trajectory. In this sense the path integral approach can be considered as 
an extension of Green's function approach \cite{K65}.  
  
Path integral technique was initially invented by Feynman for quantum mechanics and then generalised by 
Kac on various transport problems while initial ideas comes even to Wiener (see \cite{ZRS} for historical 
review). Because the technique substantially exploits the superposition concept  the 
equations under averaging has to be liner ones. It is natural for quantum mechanics which is a basically 
linear science however it is severely restrictive for dynamo theory because magnetic force begins to play a 
role at quite early stage of a dynamo driven magnetic configuration. 

Of course, people exploited path integral techniques for nonlinear dynamo studies as well (e.g. \cite{KR94})
however the results are considered as not mathematically justified just because, strictly speaking,  the Kac-Feynman formula 
is obtained for linear equations only. The point however is that the 
induction equation is a linear equation by itself and nonlinearity comes in the problem only because magnetic 
field affects motion and this effect is addressed in hydrodynamical equations. Such equations are addressed 
in mathematical studies as quasilinear ones. Path integrals method can be applied to quasilinear equations due 
to a specific trick (e.g. \cite{P91}) which exploited in   
some areas of science (in first line in financial mathematics \cite{Sh}) however the trick still deserves 
introduction in dynamo theory. This is the aim of the present paper. 

% ---------------------------- Kac-Feynman -------------------------------

\section{Kac-Feynman formula}

For the sake of consistency we consider briefly how magnetic field evolution in a given velocity field can be addressed
by path integral method. We depart from standard induction equation for magnetic field $\bf H$ in the flow $\bf v$ with magnetic diffusivity $\nu_m$

\begin{equation}
{{\partial \, {\bf H}} \over {\partial \, t}} + ({\bf v} \nabla) {\bf H} = ({\bf H} \nabla) {\bf v} - \nu_m {\rm curl} \, {\rm curl} \, {\bf H} \, .
\label{ind}
\end{equation}
For the inviscid case in Lagrangian frame of reference this equation reads

\begin{equation}
{{d \, {\bf H}} \over {d\, t}} = {\bf H} \hat A \, ,
\label{Lag}
\end{equation}
where the vector  $\bf H$ is considered as a row (not column) and matrix $\hat A$ consists on 
derivatives $\partial v_i / \partial x_j$. Variable $\bf x$ in Eq.~(\ref{Lag}) is governed by equation

\begin{equation}
{{d \, {\bf x}} \over {d \, t}} ={\bf v} \, .
\label{vel}
\end{equation}

Eq.~(\ref{Lag}) can be solved in an explicit form in terms of so-called T-exponents or, more 
mathematically, Volterra multiplicative integrals (e.g. \cite{Gant})

\begin{equation}
{\bf H} ({\bf x}(t), t) = {\bf H} \, \Pi_{s=0}^t[\hat E+ \hat A(s) d \, s],
\label{Vol}
\end{equation}
where $\hat E$ is a matrix unity. 

Eq.~(\ref{Vol}) can be extended to include the dissipative term if we add diffusion random walks $\sqrt 2 
\nu_m {\bf w}_t$ to advection velocity $\bf v$ in Eq.~(\ref{vel}). Here ${\bf w}_t$ means the standard 3D 
Wiener
random process, i.e. a process with independent increments, zero mean and correlation matrix $\delta_{ij}t
$. We have to consider a bundle of diffusion paths and summarize corresponding contributions and perform 
averaging taken over all paths (still not over realizations of velocity field). 
This is the step which requires linearity of Eq.~(\ref{ind}). From the mathematical viewpoint the last 
operation can be described as Wiener integral taken over all trajectories of fluid particle.

One has to take into account that the random walk ${\bf w}_t$ has no finite velocity
because ${\bf w}_{t+\Delta t} - {\bf w}_t \propto \sqrt t$ (Einstein relation) and we have to rewrite 
Eqs.~(\ref{vel},\ref{Vol}) in integral terms. It yields in 
 
 \begin{equation}
 {\bf H}({\bf x}, t) =M _{\bf x} {\bf H} (\mbox{\boldmath$\xi$}_{{\bf x},s}) \, \Pi_{s=0}^t[\hat E+ 
 \hat A(\mbox{\boldmath$\xi$}_{{\bf x}, s'}) d \, s'],
\label{KF}
\end{equation}
where

\begin{equation}
\mbox{\boldmath$\xi$}_{\bf{x}, s} = {\bf x} + \sqrt 2 \nu_m^{1/2} {\bf w}_t - \int_t^s {\bf v} d \, s'
\label{Ito}
\end{equation}
(so-called Ito equation; see \cite{Ito}). Eq.~(\ref{KF}) belongs to the type of expressions referred as Kac-Feynmon  
formulae. $M_{\bf x}$ mean averaging taken over trajectories obtained from Eq.~(\ref{Ito}).

The fact that Eq.~(\ref{KF}) gives solution of Eq.~(\ref{ind}) can be verified directly taken 
derivative in respect to $t$ from this function. 

Eqs.~(\ref{KF}, \ref{Ito}) do require linearity however they are exploited to get mean-field equations not 
in the full extent and it gives free space for the trick which we are going to present here.

\section{Short-correlated model for nonlinear dynamo}

To be specific, we present the trick for the so-called short-correlated model, i.e. we consider 
correlation time of turbulence or convection short enough to ignore details of the random path $\xi$ 
evolution during this time. Formally it means the following. We consider a family of random velocity fields 
${\bf v}^\Delta$ which are statistically independent and identically distributed at time intervals  $[n
\Delta, (n+1) \Delta )$.  (Note that the left boundary is included in the interval while the right one is 
excluded; it solves the problem concerning the memory of what happens just in the instants $t= n \Delta$.)
In order to avoid bulky algebra we assume that the random velocity field is 
statistically homogeneous and isotropic in space and the mean velocity vanishes (see \cite{TS10}
concerning implementation of non-zero and inhomogeneous mean velocity in the procedure and \cite{Y13} in connection with the cross helicity problem).

We are going to consider the case $\Delta \to 0$. To avoid vanishing of induction effects in this limiting 
case we have to assume that 

\begin{equation}
{\bf v}^{\Delta} \propto \Delta^{-1/2}.
\label{scale}
\end{equation} 
Scaling (\ref{scale}) looks similar to the Einstein relation $w_{t + \Delta t} - w_t \propto \sqrt{t}$. It 
means that the hydrodynamical flow in the framework of short-correlat\-ed model is supposed to be similar to the 
Brownian motion.

Then we  apply Eq.~(\ref{KF}) considering the instant $t=n \Delta$ as initial and calculate integrals 
participating in Eqs.~(\ref{KF}, \ref{Ito}) using Taylor expansions taken in respect to the parameter $
\Delta$. It means that we need to apply Eq.~(\ref{KF}) for the short time interval from $n \Delta$ till 
$(n+1) \Delta$ only rather to the whole time from 0 till $t$. 

Let us suggest that the random field ${\bf v}^{\Delta}$ depends on magnetic field on the times before the 
instant $n \Delta$ only and do not depend on magnetic field in this very interval $n \Delta \le t \le (n+1) \Delta$ . Then we can still can 
apply Eq.~(\ref{KF}) to the interval of interest in spite of the fact that the problem remains nonlinear.

This is the trick reported in the paper.

As comment on the short correlated model we note the following.
In MHD, in general, we can introduce four Green's functions:
$G_{uu}$, $G_{ub}$, $G_{bu}$, $G_{bb}$,
where $G_{fg}$ means the response of $f$ field to a unit or infinitesimal change of $g$ field.
The short correlated model in the present formulation may correspond to assuming that
the correlation times associated with $G_{bu}$ and $G_{bb}$ are much shorter than the counterpart of $G_{ub}$.

% ------------------------------ Obtaining ---------------------------------

\section{Obtaining mean-field equation}

We are now going to finalize derivation of mean-field in the framework of our model. We perform Taylor 
expansion of Eqs.~(\ref{KF}, \ref{Ito}) for $\Delta \to 0$ taking into account that

\begin{equation}
d F(w_{t}) = F' w_{dt} + {1 \over 2} F'' dt
\label{Itof}
\end{equation}
where $F$ is a smooth function (so-called Ito formula). Eq.~(\ref{Itof}) allows to restore second derivative 
term in Eq.~(\ref{ind}) from Eq.~(\ref{KF}).

Eq.~(\ref{Ito}) yields in

\begin{equation}
\mbox{\boldmath$\xi$}_{i, \Delta} - x_i = - v_i (n \Delta, x) \Delta + \sqrt 2 \nu_m^{1/2} w_{i, \Delta} - {{\partial v_i} \over {\partial x_j}} \int_0^\Delta w_{t, j} dt + {1 \over 2} v_j {{\partial v_i} \over {\partial x_j}} \dots
\label{expIt}
\end{equation}
where dots stand for the terms smaller rather $\Delta$ (remember that the second term in the right side is 
of order $\Delta$ because of Eq.~(\ref{scale}) while the multiplicative integral in (\ref{KF}) can be presented as

\begin{eqnarray}
\Pi_{s=0}^t[\hat E+ 
\nonumber
 \hat A(\mbox{\boldmath$\xi$}_{\bf x}, s') d \, s'] = \delta_{ij} - \int_0^\Delta {{\partial v_i (t-s,\mbox{\boldmath$\xi$}_s)} \over {\partial x_j}} + \\
 + \int_o^\Delta {{\partial v_i (t-\sigma, \mbox{\boldmath$\xi$}_\sigma) } \over {\partial x_l}} \int_0^\Delta {{\partial v_l (t-s, \mbox{\boldmath$\xi$}_s) } \over {\partial x_j}} ds \, d\sigma + \dots\, .
\label{mulin}
\end{eqnarray}
Taking into account that

\begin{equation}
{{\partial v_i (t-s, \mbox{\boldmath$\xi$}_s)} \over {\partial x_j}} = {{\partial v_i (t-s, x)} \over {\partial x_j}} + {{\partial^2 v_i} \over {\partial x_j \partial x_k}} (\mbox{\boldmath$\xi$}_{s, k} - x_k)+\dots \, ,
\label{approac}
\end{equation}
collecting terms in Eq.~(\ref{KF}) and performing averaging following \cite{Metal83} we obtain mean-field equation for ${\bf B} = <{\bf H}>$ as follows

\begin{equation}
{{\partial {\bf B}} \over {\partial t}} = {\rm curl} \, \alpha {\bf B} - \beta \, {\rm curl} \, {\rm curl} \,  {\bf B} \, ,
\label{SKR}
\end{equation}
where

\begin{equation}
\alpha = \lim_{\Delta \to 0} \Delta {{<{\bf v} {\rm curl} \, {\bf v}>} \over 3}
\label{alp}
\end{equation} 
and

\begin{equation}
\beta = \lim_{\Delta \to 0}\Delta {{<{\bf v}^2>} \over 3}.
\label{td}
\end{equation}

Physical meaning of the limiting procedure $\Delta \to 0$ is that we are not interesting in what happens 
in timescales shorter than $\Delta$ which plays a role of the memory time $\tau$. Standard expressions of the 
mixing length theory for $\alpha$ and $\beta$ 
follow from (\ref{SKR}) and (\ref{td}) using $l=v \Delta$ as a spatial scale of the flow.

% ------------------------------ Conclusions ---------------------------------

\section{Conclusions and discussion}

Using the path integral method in the framework of above described model we obtained mean-field dynamo 
equation (\ref{SKR}) for nonlinear dynamo. The form of this equation is just the same as conventional 
mean-field equation obtained by Steenbeck, Krause and R\"adler (see \cite{KR}). The important difference 
from the classical mean-field equation is that the averaging in Eqs.~(\ref{alp}, \ref{td}) is taken over the 
ensemble 
of velocity field realisations affected by magnetic force while classical mean-field equation presumes that 
this ensemble as well as the quantities (\ref{alp}, \ref{td}) are given in advance. 
This departure from the kinematic approach must be the most important point in the present formulation of 
nonlinear dynamo.

In this sense 
Eq.~(\ref{SKR}) is not a close linear equation however should be combined with equations for calculation of 
the quantities 
(\ref{alp}, \ref{td}). Of course, these equations have to be obtained from some arguments external in 
respect to 
the above consideration. As a possible source for a procedure for calculation of quantities from 
Eqs.~(\ref{alp}, \ref{td}) we could suggest 
the technique of shell models \cite{Petal13} which looks at the present stage of turbulent studies a most 
promising technique for quantification of the locally homogeneous and isotropic turbulence for very high 
hydrodynamic and magnetic Reynolds numbers and a generalization for anisotropic case looks debatable.

The above demonstrated evaluation of the mean-field equation becomes possible because we presume that 
nonlinear magnetic force action happens before the renovation instant $t=n \Delta$ while corresponding 
induction effect acts from the instant $t= n\Delta$ till the instant $t= (n+1) \Delta$. As other
assumptions which allow to get closed mean-field equations, this assumption is a simplification and its 
applicability is limited.   

An important feature of Eq.~(\ref{SKR}) is that it yields in the mean-field equation which does not include 
the concept of magnetic helicity as a 
quantity important for nonlinear mean-field dynamo suppression. As a matter of fact, magnetic helicity does 
not participate in this equation directly and can participate via Eqs.~(\ref{alp}, \ref{td}) only. Maybe, this 
is a shortcoming of the model of the flow exploited here. If it is the matter of fact, it means that the key 
issue 
for arising of magnetic helicity term in nonlinear dynamo model is the absence of any time lag between 
induction effect and magnetic force action. Such lag is postulated in our model. 
In the usual homogeneous turbulence theory, it is often considered that the  helicity introduces a 
timescale other than the so-called eddy-turnover time. For instance, the bottleneck effect (energy pile up at 
small scales) in helical turbulence is often attributed to the timescale change due to turbulence at the small 
scales. The time lag argument may be able to consider with this respect.

In any case, the final decision which parametrisation for nonlinear mean-field dynamo action is more 
realistic, has to come from experiences with modelling of particular natural (or, in perspective, 
laboratory) dynamos.

% ------------------------------------ Acck --------------------------------
\section*{Acknowledgements}

The paper is prepared in the framework of the Russia -- Japan Bilateral Project (RFBR -- JSPS) 2016-2017, RFBR grant number 16-52-50077.
 
% ------------------------------------ Bib --------------------------------

\clearpage

% ------------------------------------------------------------------------


\begin{thebibliography}{99}

\bibitem[Gantmacher, 1959]{Gant}
\textsc{Gantmacher F.R.} 1959
Applications of the theory of matrices.
\textit{Interscience., NY}.

\bibitem[Ito, 1946]{Ito}
\textsc{Ito, K.} 1946
On stochastic integral equation.
\textit{Proc. Japan. Acad.}, \textbf{1}, 32--35.


\bibitem[Kleeorin \& Rogachevsky, 1994]{KR94}
\textsc{Kleeorin, N., \& Rogachevskii, I.} 1994
Nonlinear theory of magnetic fluctuations in random flow: The Hall effect.
\textit{Phys. Rev. E.}, \textbf{50}, 493--501.

\bibitem[Kraichnan, 1965]{K65}
\textsc{Kraichnan, R.-H.} 1965
Lagrangian-history closure approximation for turbulence.
\textit{Phys. Fluids}, \textbf{8} (1965), 575--598.


\bibitem[Krause \& R\"adler, 1980]{KR}
\textsc{Krause, F. \& R\"adler, K.-H.} 1980
Mean-field magnetohydrodynamics and dynamo theory.
\textit{Oxford: Pergamon}.

\bibitem[Molchanov \etal, 1983]{Metal83}
\textsc{Molchanov, S.A., Ruzmaikin, A.A., \& Sokoloff, D.D.} 1983
Equation of dynamo in random velocity field with
short correlation time.
\textit{Magnetohydrodynamics}, \textbf{19}, 402--407.

\bibitem[Parker, 1955]{P55}
\textsc{Parker, E.N.} 1955
Hydromagnetic dynamo models.
\textit{Astrophys. J.} \textbf{122}, 293--314.


\bibitem[Peng, 1991]{P91}
\textsc{Peng, S.} 1991
Probabilistic interpretation for systems of quasilinear parabolic partial differential equations.
\textit{Stochastics and stochastic reports}, \textbf{37} (1991), 61--74.



\bibitem[Plunian \etal, 2013]{Petal13}
\textsc{Plunian, F., Stepanov, R., \& Frick, P.} 2013
Shell models of magnetohydrodynamic turbulence.
\textit{Physics Reports}, \textbf{523}, 1--60.



\bibitem[Shiryaev, 1999]{Sh}
\textsc{Shiryaev, A.N.} 1999
Essentials of Stochastic Finance: Facts, Models, Theory.
\textit{World Sci., Singapore}.

\bibitem[Tomon \& Sokoloff, 2010]{TS10}
\textsc{Tomin, D. \& Sokoloff, D.} 2010
Dynamo in fluctuating ABC flow.
\textit{Geophys. \& Astrophys. Fluid Dyn.}, \textbf{104}, 183-188.

\bibitem[Yokoi, 2013]{Y13}
\textsc{Yokoi, N} 2013
Cross helicity and related dynamo.
\textit{Gepphys. \& Astrophys. Fluid Dyn.}, \textbf{107}, 114-184.

\bibitem[Zeldovich \etal, 1990]{ZRS}
\textsc{Zeldovich, Ya.B., Ruzmaikin, A.A., \& Sokoloff, D.D.} 1990
The almighty chance.
\textit{World Sci., Singapore}.





\end{thebibliography}
\end{document}